%% file: main.tex
\pdfoutput=1

\documentclass[11pt]{article}

\usepackage{acl}

\usepackage{times}
\usepackage{latexsym}

\usepackage[T1]{fontenc}

\usepackage[utf8]{inputenc}

\usepackage{microtype}

\usepackage{amsfonts}
\usepackage{amsmath}
\usepackage{multirow,makecell}
\usepackage{booktabs}
\usepackage{graphicx,subfigure}
\newcommand{\hide}[1]{} 

\def\by{{\mathbf{y}}}
\def\bz{{\mathbf{z}}}

\def\be{{\mathbf{e}}}

\def\bp{{\mathbf{p}}}

\def\cV{{\mathcal{V}}}

\def\RR{{\mathbb{R}}}

\def\II{{\mathbb{I}}}
\def\TARGET{{\texttt{<extra\_id\_10>}}}
\def\TRUE{{\texttt{true}}}
\def\FALSE{{\texttt{false}}}

\def\enc{\text{Enc}}
\def\dec{\text{Dec}}
\def\dense{\text{Dense}}
\def\softmax{\text{Softmax}}

\title{RankT5: Fine-Tuning T5 for Text Ranking with Ranking Losses}

\author{
    Honglei Zhuang, Zhen Qin, Rolf Jagerman, Kai Hui, Ji Ma, Jing Lu, Jianmo Ni, \\
    \bf{Xuanhui Wang \and Michael Berdersky} \\
    
    Google Research \\

    \texttt{\{hlz,zhenqin,jagerman,kaihuibj,maji,ljwinnie,jianmon,}\\
    \texttt{xuanhui,bemike\}@google.com}     
}

\begin{document}

\maketitle
\input{abstract.tex}
\input{intro.tex}

\input{related.tex}

\input{prelim.tex}

\input{model.tex}

\input{exp.tex}

\input{results.tex}

\input{conclusion.tex}

\bibliography{references}
\newpage

\input{appendix.tex}
\end{document}

%% file: abstract.tex
\begin{abstract}

Recently, substantial progress has been made in text ranking based on pretrained language models such as BERT. However, there are limited studies on how to leverage more powerful sequence-to-sequence models such as T5.
Existing attempts usually formulate text ranking as classification and rely on postprocessing to obtain a ranked list.
In this paper, we propose RankT5 and study two T5-based ranking model structures, an encoder-decoder and an encoder-only one, so that they not only can directly output ranking scores for each query-document pair, but also can be fine-tuned with ``pairwise'' or ``listwise'' ranking losses to optimize ranking performances.
Our experiments show that the proposed models with ranking losses can achieve substantial ranking performance gains on different public text ranking data sets.
Moreover, when fine-tuned with listwise ranking losses, the ranking model appears to have better zero-shot ranking performance on out-of-domain data sets compared to the model fine-tuned with classification losses.

\end{abstract} 

%% file: intro.tex
\section{Introduction}
\label{sec:intro}

Text ranking refers to the task of ranking a set of textual documents based on their relevance to a given query or context.
Learning a text ranking model is a fundamental component of countless real world applications such as search and question answering.
Earliest explorations~\cite{liu2009ltr} mostly rely on handcrafted numerical features to represent each query-document pair~\cite{qin2013letor,chapelle2011yahoo} and put more emphasis on the learning algorithms such as ranking losses~\cite{qin2021AreNeuralRankers}.
Progress on pretrained language models in the past few years~\cite{devlin2019bert} and the release of large-scale public data sets~\cite{msmarco,nq} enable a series of work~\cite{lin2020pretrained,monobert,tfrbert} on text ranking models which directly encode textual query and document using pretrained language models, noticeably BERT~\cite{devlin2019bert}. 

Recently, pretrained language models such as T5~\cite{raffel2020exploring} and GPT3~\cite{brown2020language} have shown superior performance in various NLP tasks including sentiment analysis, coreference resolution, and translation. 
Such models often have much larger size available than previous models such as BERT~\cite{devlin2019bert} to store more hidden knowledge.
They also mostly have a sequence-to-sequence interface to unify different NLP tasks from classification to text generation. 

While BERT-based models have been well explored for text ranking~\cite{lin2020pretrained, monobert, tfrbert}, how to leverage T5 for text ranking is still under-explored and challenging.
First, while many classification and text generation tasks fit into the sequence-to-sequence framework, it is more tricky for text ranking tasks: 
a text ranking model is often expected to output a numerical ranking score $\hat{y} \in \RR$ for each query-document pair. 
Second, it is important to train a text ranking model with ranking losses to optimize its ranking performance, where the losses take into account the ranking scores from multiple documents for each query. 
This is different from the typical T5 fine-tuning strategy where the objective is often formulated into a text generation loss for each single input sequence independently. 

A typical approach to use T5 for text ranking is to convert the problem into a token generation problem. 
For example, ~\citet{nogueira2020T5ranking} fine-tune the T5 model to predict a ``\TRUE'' or ``\FALSE'' token for a relevant or irrelevant query-document pair and then use a postprocessing step during inference to derive ranking scores to rank candidate documents. 
Such an approach can be considered a ``pointwise'' classification formulation. 
How to extend this approach to fine-tune T5 with ranking losses is unclear.

In this paper,
we propose RankT5 with the goal to support text ranking more natively with T5 by outputting real numbers, instead of text tokens. 
We first adapt the encoder-decoder structure for this goal. 
In addition, we also propose an alternative structure which omits the decoder from T5 and outputs real numbers based on the encoder, called the encoder-only structure.
These two structure variants allow us to fine-tune T5 with various ranking losses to directly optimize ranking performance.

Experiments on MS MARCO and Natural Question data sets show that our RankT5 models fine-tuned with specialized ranking losses can significantly outperform other T5 ranking models fine-tuned with classification losses and previously proposed T5 adaptations for ranking~\cite{nogueira2020T5ranking}. 
We also discover that models fine-tuned with some ranking losses tend to have better zero-shot performance than models fine-tuned with classification losses.

%% file: related.tex
\section{Related Work}
\label{sec:related}

Leveraging pretrained language models for text ranking tasks is becoming the state-of-the-art~\cite{lin2020pretrained}. 
We review recent literature on utilizing such models for text ranking tasks. 

\paragraph{Model structure.}
Pretrained language models accept a text sequence as input.
A typical model structure design is the cross-attention model structure, 
where a query and a candidate document is concatenated into a sequence and fed into the model. 
It allows the attention mechanism to fully capture the query-document interaction.
This model structure has been explored for BERT-like encoder-only model~\cite{tfrbert,monobert,gao2021rethink}. 
This is also explored in the T5-like model~\cite{nogueira2020T5ranking,ju2021text}, but the model is not directly fine-tuned with ranking losses for the optimal ranking performance. 
Instead, the model is fine-tuned to generate ``\TRUE'' or ``\FALSE'' tokens. The ranking scores are later derived in postprocessing from the predicted logits. 
There are also model structures that take a triple with one query and two documents as input~\cite{pradeep2021expando}, but these models can only be deployed in the \emph{late ranking} stage to work with tens of documents because they require scoring all possible input document pairs, which is not scalable. Our contributions are in the \emph{early ranking} stage which scores thousands of input documents; they are thus complementary with this work. 

There are also works exploring ranking models based on generative likelihood from language models. 
For example, existing works in~\cite{Zhuang2021-dn,zhuang2021deep,sachan2022improving,dos2020beyond} take a candidate document as input and estimates the likelihood of generating the query. Then they rank documents based on the query-likelihood or use it as an auxiliary task~\cite{ju2021text}.
These models convert ranking into a generation task in order to leverage generative models such as T5, instead of directly enabling T5 to output numerical scores.

There are a few other model structures proposed for \textit{retrieval} tasks,
where the model needs to score hundreds of millions of documents in the entire corpus almost instantly for each arriving query. 
These model structures emphasize more on model efficiency. 
Typical examples for BERT-like models are
dual-encoder models~\cite{karpukhin2020dense} and 
encoders with late interaction~\cite{khattab2020colbert,gao2020modularized,macavaney2020efficient},
where the model computation of queries and documents are decomposed to allow precomputations for documents in the corpus. 
Some other works also utilize the predictive likelihood from generative T5-like models for retrieval. 
Examples include using the likelihood of generating queries from documents~\cite{ed2lm2022} or using the likelihood of generative document IDs from queries~\cite{tay2022transformer}.
However, our paper focuses on the ranking task, where a small set of candidates documents are already retrieved for each query. 
This allows us to focus on cross-attention model structure and push for better ranking performance. 

\paragraph{Fine-tuning with ranking losses.}

Early explorations~\cite{monobert,nogueira2020T5ranking} of applying pretrained language models on the document reranking task mainly use ``pointwise'' losses, where the loss is calculated for each query-document pair independently. 
There are some recent works that use a ``listwise'' loss~\cite{tfrbert,gao2021rethink,ren2021rocketqav2} which takes one positive and multiple negative documents for each query and calculates the softmax loss over their ranking scores.
But they only fine-tune BERT, RoBERTa etc.
There is no existing work fine-tuning sequence-to-sequence models like T5 with \emph{ranking losses}.

Some retriever models are fine-tuned using pairwise loss or softmax loss~\cite{xiong2020approximate,karpukhin2020dense,lu2021multi}. 
However, in this work we only focus on reranking models.

%% file: prelim.tex
\section{Preliminaries}
\label{sec:prelim}

\paragraph{Problem definition.}
We provide the formalized definition of a ranking task.
For each query $q_i$, a list of candidate documents $D_i = (d_{i1}, \ldots, d_{im})$ are provided, which are usually the output from a retriever. 
The relevance labels of candidate document with regard to the query are represented as $\by_i = (y_{i1}, \ldots, y_{im})$ where $y_{ij} \geq 0$.

The objective is to train a ranking model $f$ which takes a query-document pair as input and outputs a ranking score $\hat{y}_{ij} = f(q_i, d_{ij}) \in \mathbb{R}$.
We aim to optimize the ranking metrics after we sort the documents $D_i$ for each query $q_i$ based on their ranking scores.  

In the ranking task, the number of documents in $D_i$ given for a query is usually small and $f$ can be a high-capacity model such as a cross-attention model. 
On the contrary, a retrieval task~\cite{qu2021rocketqa} virtually needs to score all the documents in a corpus with millions of documents.
In this study we only focus on the ranking task.

\paragraph{T5.}
T5~\cite{raffel2020exploring} is a text-to-text pretrained generative language model with an encoder-decoder structure based on transformers. 
The original T5 takes a piece of text as input and outputs a sequence of text tokens in an autoregressive manner. 
We can also obtain the output likelihood of each token in the vocabulary $\cV$ given the input text. 

More formally, we denote the input to the T5 encoder as a text sequence $s = [w_1, \ldots, w_l]$ and the previously 
generated tokens from the T5 decoder as $t_{1:k-1}$ during the autoregressive decoding process. 
We formalize the T5 model structure as:
\begin{align}
    \bp_k &= \text{T5}(s, t_{1:k-1})  \nonumber \\ 
          & = \softmax(\dense(\dec(\enc(s), t_{1:k-1}))) \nonumber
\end{align}
where the output is a vector with the length of the vocabulary size $\bp_k \in \RR^{|\cV|}$, representing the predictive probability of each token in the vocabulary being generated at the $k$-th position.

The following components are included in the T5 model structure: $\enc(\cdot)$ is the encoder of T5 which outputs a sequence of embedding vectors; $\dec(\cdot, \cdot)$ is the decoder of T5 which takes both the encoder outputs and the previously generated tokens from the decoder as inputs and outputs a single embedding vector; $\dense(\cdot)$ is a dense layer that projects the embedding vector into a unnormalized logits vector with length $|\cV|$; and $\softmax(\cdot)$ normalizes the vector into probabilities.

%% file: model.tex
\section{RankT5 Modeling}
\label{sec:model}

\begin{figure*}[t]
  \centering
  \subfigure[monoT5~\cite{nogueira2020T5ranking}]{
    \label{subfig:model_monot5}
    \includegraphics[width=0.3\textwidth]{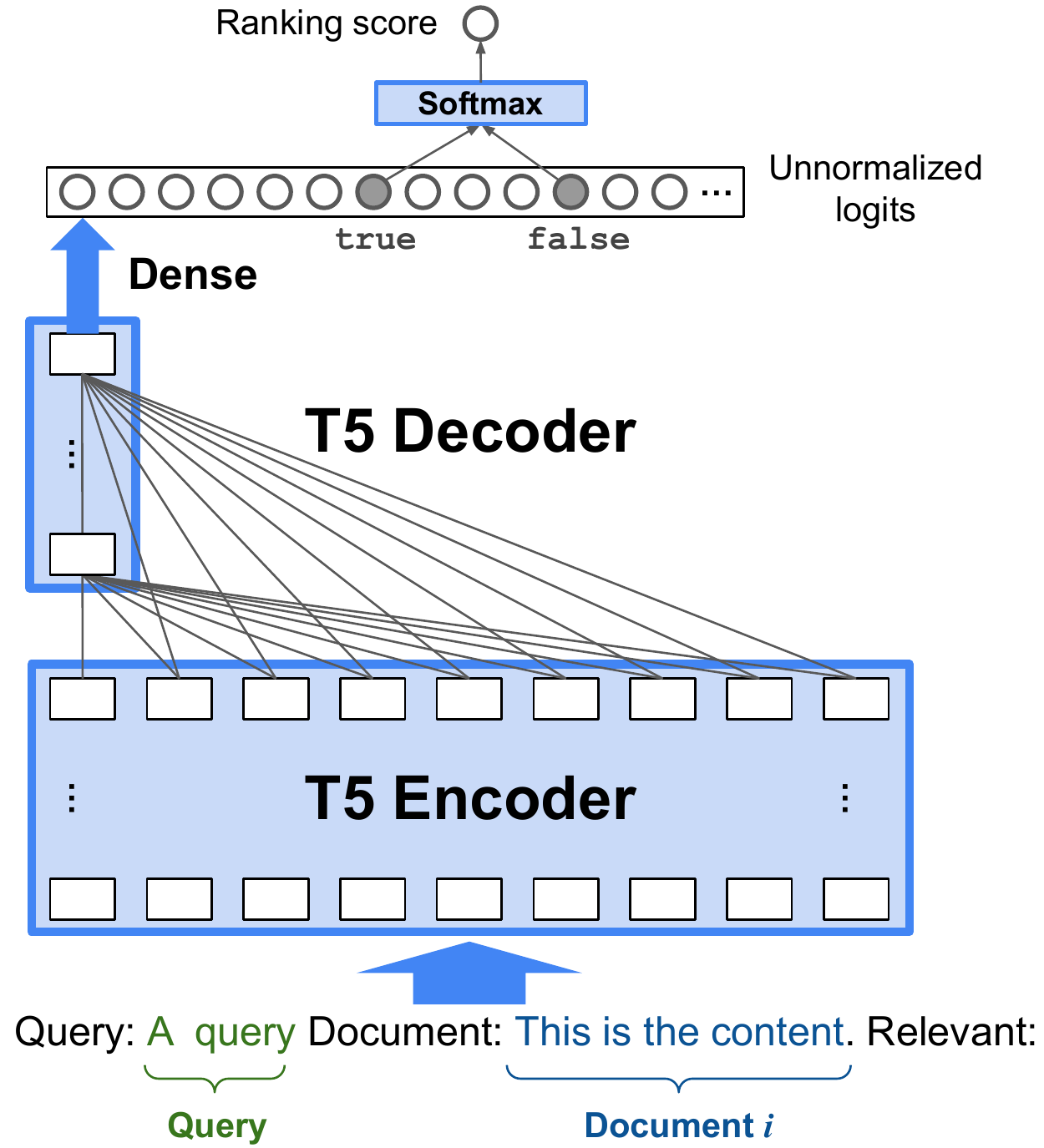}
  }
  \subfigure[RankT5 (Encoder-Decoder)]{
    \label{subfig:model_encdec}
    \includegraphics[width=0.3\textwidth]{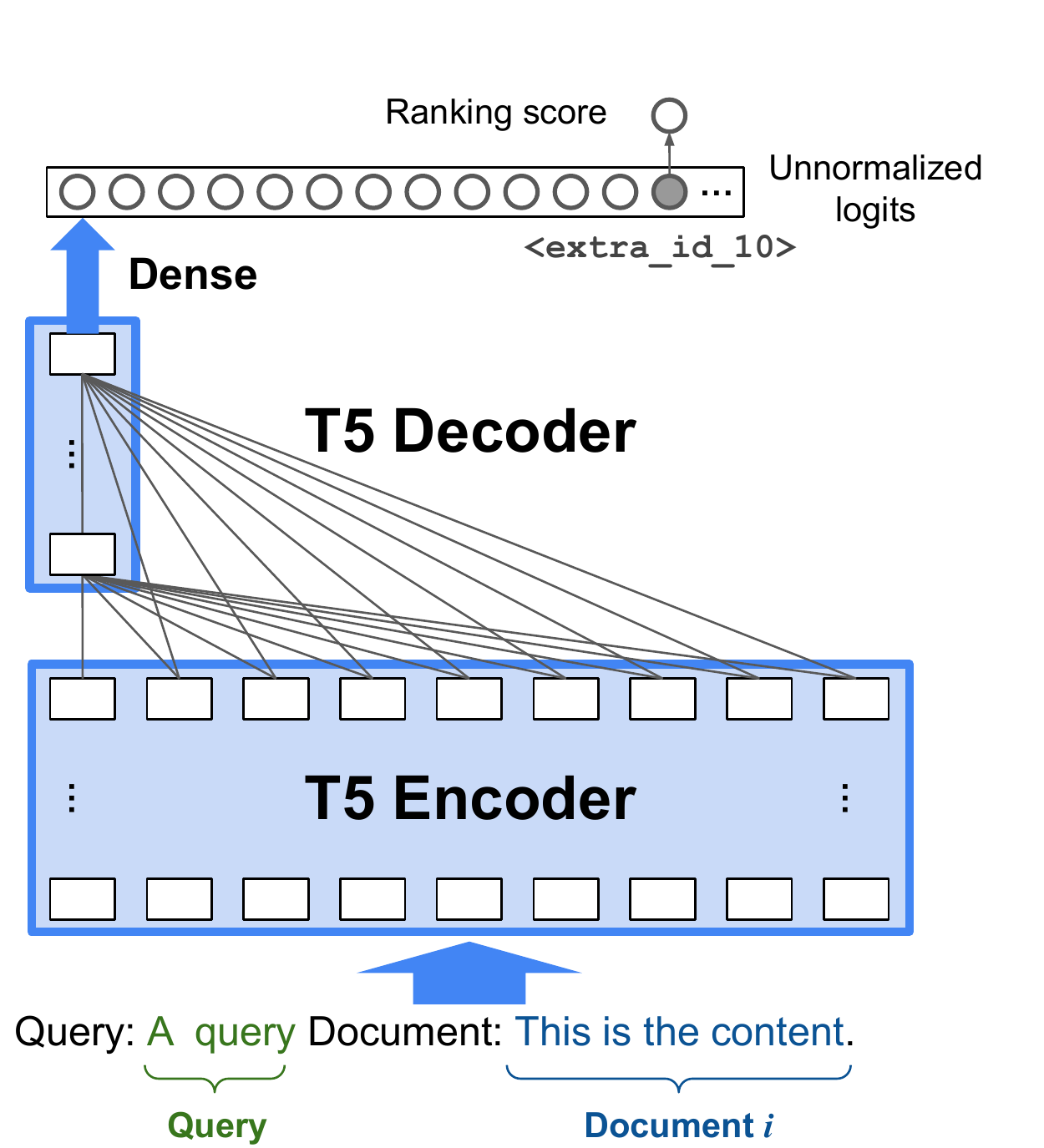}
  }
  \subfigure[RankT5 (Encoder-Only)]{
    \label{subfig:model_enc}
    \includegraphics[width=0.3\textwidth]{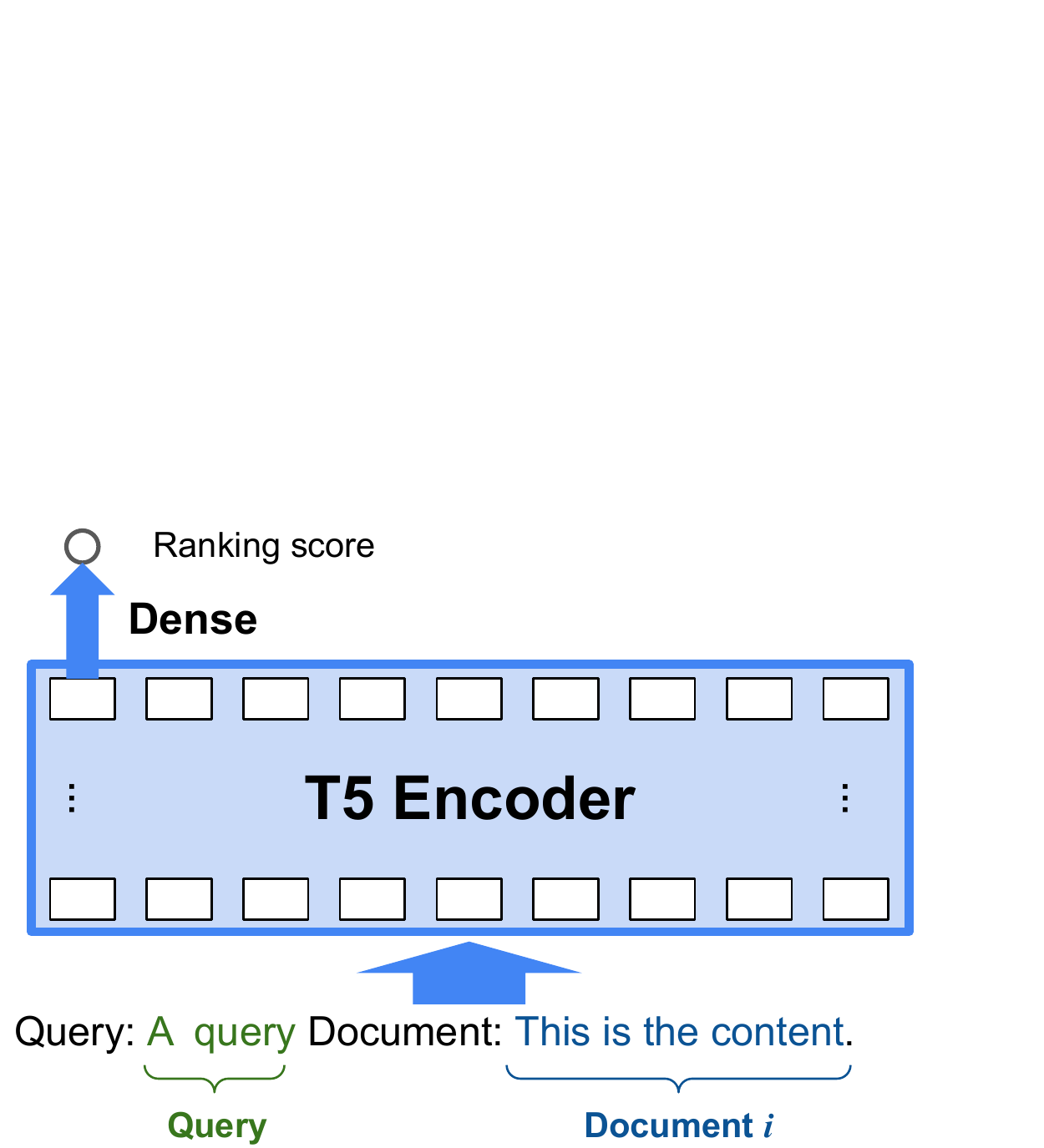}
  }
  \caption{
   Illustration of the T5 ranking model structure from prior work and the two variants of our proposed RankT5 model. The differences in training mechanism is not reflected in this illustration.
  }
  \label{fig:model_structure}
\end{figure*}

\subsection{Model structure}
\label{subsec:model_structure}

Applying T5 on generative tasks (e.g., summarization, translation) is straightforward due to the sequence-to-sequence nature of these tasks.  
T5 can also be straightforwardly adapted for classification tasks such as sentiment analysis or textual entailment recognition~\cite{raffel2020exploring}. 
A commonality of these tasks is that the output are text tokens or sequences of text tokens. 
How to use T5 in a ranking model has not been studied widely because the output are not tokens, but numerical scores used for sorting.

\paragraph{Existing work.} \citet{nogueira2020T5ranking} cast text ranking as classification to fit the T5 model structure.  
They concatenate each query-document pair into an input sequence and use ``\TRUE'' and ``\FALSE'' as target tokens to indicate whether they are relevant or not; the T5 model is then fine-tuned as a text generation task.
After the model is fine-tuned, the ranking scores are obtained from the logits of ``\TRUE'' and ``\FALSE'' by $\hat{y}_{ij} = e^{\bz_{(\TRUE)}} / (e^{\bz_{(\TRUE)}} + e^{\bz_{(\FALSE)}})$, where $\bz_{(w)}$ represents the logits of $w$ in the vocabulary $\cV$.
Figure~\ref{subfig:model_monot5} shows this model structure.

This existing method only obtains the numerical ranking score as a postprocessing step after the fine-tuning is done.
Therefore, the model is not directly fine-tuned as a ranking model but as a text generation or classification model, which may not optimize its ranking performance. 
We propose to directly obtain the numerical ranking score as the model output. 
Specifically, we implement two variants based on pretrained T5 models: an encoder-decoder and an encoder-only model.

\paragraph{Input sequence.}
For each candidate document $d_{ij}$ and its query $q_i$, we concatenate them with prefix ``Document:'' and ``Query:'' respectively to construct the input text $s_{ij}$:
\begin{align}
    s_{ij} = \text{Query:}~q_i~\text{Document:}~d_{ij} \nonumber
\end{align}
The construction of input sequence is similar to~\citet{nogueira2020T5ranking} except that we do not include the ``Relevant:'' postfix. The postfix does not affect the results in our experiments (Appendix~\ref{sec:target_token}).

\paragraph{Encoder-decoder (EncDec).}
This model variant is a simple adaptation of the T5 model by using the first output token of the decoder. In this variant, we feed the input into a T5 model and obtain the unnormalized logits $\bz$ over the entire vocabulary:
\begin{align}
  \bz = \dense(\dec(\enc(s_{ij})))
\end{align}
Notice that we omit the softmax layer over vocabulary so that the elements in $\bz$ can be arbitrary real numbers. 
We also simplify the decoder representation $\dec(\cdot, \cdot)$ into $\dec(\cdot)$ which only takes the output from the encoder as its input, since we do not generate other tokens from the decoder.

We then specify a special unused token in T5 vocabulary ``\TARGET'' and take its corresponding unnormalized logits as the ranking score: 
\begin{align}
 \hat{y}_{ij} = \bz_{(\TARGET)}
\end{align}
where we use the notation $\bz_{(w)}$ to represent the logits corresponding to the token $w \in \cV$.
The special token can be any other unused token in the vocabulary. 
An illustration of this model structure can be found in Figure~\ref{subfig:model_encdec}. 

While an experiment in~\cite{nogueira2020T5ranking} shows that directly using a single token's logit (in their case $\hat{y}_{ij} = \bz_{(\TRUE)}$) works poorly,
notice that a key difference in this work is that the training loss is a ranking loss directly defined on top of the ranking score $\hat{y}_{ij}$, instead of using the original text generation loss for T5 fine-tuning. 
We will provide more details in Section~\ref{subsec:training}.

\paragraph{Encoder-only (Enc).}
Since we do not need to perform autoregressive decoding, the decoder of T5 could be unnecessary. 
Therefore, we also try an encoder-only variation.

We construct the input text $s_{ij}$ in a similar manner and feed them into the T5 encoder. 
The output of the encoder $\enc(s_{ij})$ is a sequence of embedding vectors $[\be_1, \cdots, \be_l]$ with the same sequence length of the input sequence. 
\begin{align}
 [\be_1, \cdots, \be_l] = \enc(s_{ij})
\end{align}
We use a pooling layer $\text{Pool}(\cdot)$ (e.g., using first-token's embedding or mean pooling) to aggregate them into a single embedding vector. 
Then, we apply the dense layer which directly projects the embedding vector to the ranking score $\hat{y}_{ij}$. 
To formally summarize the encoder-only model:
\begin{align}
  \hat{y}_{ij} = \dense(\text{Pool}(\enc(s_{ij})))
\end{align}

Figure~\ref{subfig:model_enc} summarizes the proposed model structure. 
The T5 encoder-only model structure is very similar to BERT.
The major difference is that the initial model parameter checkpoint will be from T5 instead of BERT.

\paragraph{Remarks.}
We propose to try the encoder-only structure because it is unclear how important the decoder of T5 is for text ranking tasks, especially when there is abundant labeled data to fine-tune the model. 
The decoder model structure itself contains attention layers across all the embedding vectors from the output of the encoder, essentially serving as a complicated pooling layer with a lot more parameters. 
By replacing the decoder with a simple pooling layer in the encoder-only structure, we remove both the potential hidden knowledge learned by the decoder during the pretraining stage of T5, as well as the extra modeling capability of pooling. 
Comparing the performance between these two model structures can provide some insights on whether they are essential for text ranking tasks.

\subsection{Training}
\label{subsec:training}

The adapted model structure enables the model to directly output the ranking score for each query-document pair. 
We can thereby adopt the training manner similar to learning-to-rank models.

For each query $q_i$ and a list of its candidate documents $D_i$, we obtain the list of predicted ranking scores $\hat{\by}_i$ by applying the model on each query-document pair $(q_i, d_{ij})$ where $d_{ij} \in D_i$. 
Then we can train the model by optimizing a ranking-based training loss function $\ell(\by_i, \hat{\by}_i)$ which is defined on the list of predicted scores $\hat{\by}_i$ and the list of ground-truth relevance labels $\by_i$. We study the following ranking losses:

\paragraph{Pointwise cross entropy (PointCE).}
We first try the pointwise cross-entropy loss function which is calculated based on each query-document pair independently.
\begin{align}
  & \ell_{\text{PointCE}}(\by_i, \hat{\by}_i)  \nonumber \\
= &-\sum_{j | y_{ij} = 1} \log(\sigma(\hat{y}_{ij})) -\sum_{j | y_{ij} = 0} \log (1 - \sigma(\hat{y}_{ij}))  \nonumber
\end{align}
where $\sigma(x) = \frac{1}{1 + e^{-x}}$ is the logistic function. 

The loss basically converts the ranking problem into a binary classification problem for each query-document pair.
This loss is only applicable when the relevance labels are binary.

\paragraph{Pairwise logistic (Pair).}
Instead of regarding each query-document pair independently, we can also focus on pairwise comparison.
For example, we can train the model using a pairwise logistic ranking loss~\cite{burges2005learning}:
\begin{align}
    \ell_{\text{Pair}}(\by_i, \hat{\by}_i) = &\sum_{j=1}^{m} \sum_{j'=1}^{m} \II_{y_{ij} > y_{ij'}} \log \big(1 + e^{\hat{y}_{ij'} - \hat{y}_{ij}}\big)   \nonumber
\end{align}
where the ranking problem is converted into a binary classification problem on the order of each candidate document pair in the same list with different relevance labels.

\paragraph{Listwise softmax cross entropy (Softmax).}
We can also define a listwise softmax cross entropy loss~\cite{bruch2019softmax}, which takes the entire list into account.
It can also be considered as a simple version of ListNet~\cite{cao2007listnet}.
\begin{align}
    \ell_{\text{Softmax}}(\by_i, \hat{\by}_i) = -\sum_{j=1}^{m} y_{ij} \log\Big(\frac{e^{\hat{y}_{ij}}}{\sum_{j'} e^{\hat{y}_{ij'}}}\Big)   \nonumber
\end{align}

\paragraph{Listwise poly-1 softmax cross entropy (Poly1).}

We also try a recently proposed extended version of the softmax cross-entropy loss called PolyLoss~\cite{leng2022polyloss}.
The idea is to adjust the weights of polynomial terms in the Taylor expansion of a softmax cross entropy loss.
A simplified version only adjusted the first polynomial term: 
\begin{align}
    \ell_{\text{Poly1}}(\by_i, \hat{\by}_i) &= \ell_{\text{Softmax}}(\by_i, \hat{\by}_i) \nonumber \\
    &+ \sum_{j=1}^m \epsilon\cdot y_{ij'} \Big(1 -  \frac{e^{\hat{y}_{ij}}}{\sum_{j'} e^{\hat{y}_{ij'}}}\Big)   \nonumber
\end{align}
where $\epsilon$ is a parameter specified by users, representing how much extra coefficient to be placed for the first polynomial term.

\paragraph{Discussion.}
In the previous work~\cite{nogueira2020T5ranking} the monoT5 ranker is still fine-tuned by a text generation loss.
The generation loss for each token is defined on a softmax function over the entire vocabulary, and two separate target tokens ``$\TRUE$'' and ``$\FALSE$'' are used during the fine-tuning. 
Hence only using the logits of single target token ``$\TRUE$'' as ranking scores does not align with the fine-tuning set up and cannot achieves reasonable results.
If they apply the constraints during fine-tuning that the logits for the ``$\FALSE$'' target token are always 0 and all the other tokens are excluded, 
they should be able to only use the logits of ``$\TRUE$'' as ranking scores.
When there are abundant training data, it should be practically equivalent to our RankT5 with the encoder-decoder structure fine-tuned with the PointCE loss.
With these commonalities, we expect our RankT5-EncDec model performs no worse than monoT5 when fine-tuned with the PointCE loss.


%% file: exp.tex
\section{Experiment Setup}
\label{sec:exp}

\subsection{Data sets}

\paragraph{MS MARCO.}
We use the MS MARCO passage ranking data set~\cite{msmarco}.
The data set contains around 530,000 queries in the ``train'' partition and around 6,800 queries in the ``dev'' partition. 
The candidates are from a corpus with more than 8.8 million passages.
For each query, relevant passages are labeled as 1 and others are labeled as 0.
We use a dual-encoder retriever~\cite{ni2021large} fine-tuned on MS MARCO to retrieve the top-1000 passages for each query in both the ``train'' and ``dev'' partitions as the candidate documents.

Notice that in this paper, the term ``document'' is used interchangeably with ``passage''. 
We do \emph{not} focus on long document ranking tasks such as the MS MARCO document ranking task. 

\paragraph{Natural Questions (NQ).}
We use the Natural Questions data set~\cite{nq}, where there are more than 50,000 queries in the ``train'' partition and 8,000 in the ``dev'' partition. 
We adopt the preprocessing setup similar to~\citet{karpukhin2020dense} to construct the corpus of passages. 
Similar to MS MARCO, binary relevance labels are provided. 
We use a dual-encoder retriever~\cite{lu2021multi} fine-tuned on NQ to retrieve the top-1000 passages for each query.

\paragraph{Training data construction.}
We construct the training data by first selecting a document with label 1 for each query and then uniformly randomly sampling $(m - 1)$ documents from the top-1000 retrieved documents with labels 0. 
Due to the size of T5 models and hardware constraints, the maximum $m$ we can support is 36 with the batch size 32 while fine-tuning T5-Large models. 
We will show how $m$ impacts the ranking performance in our experimental results. 

For models with pointwise training losses, we upsample documents with label 1 in each query to the same number as documents with label 0 in order to achieve the optimal performance.

\subsection{Evaluation}
We evaluate the performance on the ``dev'' partition on both data sets. 
We perform model inference on the top-1000 documents retrieved by the dual-encoder retriever of each data set respectively. 
Please note that our evaluations are applied on top 1000 documents while our training only takes a sample of 36 documents for each query.

We evaluate the performance by Mean Reciprocal Rank (MRR@10)~\cite{voorhees1999trec} and Normalized Discounted Cumulative Gain (NDCG@$k$)~\cite{jarvelin2002cumulated} where $k=5, 10$.

\subsection{Parameter configurations}

We initialize the ranking model with pretrained T5 checkpoint.
If not specified, the default size we use in our experiments is T5-Large. 
For the pooling layer in RankT5-Enc, we just follow BERT~\cite{devlin2019bert} and take the embedding vector of the first token.
These results do not differ when using mean pooling (Appendix~\ref{sec:enc_pooling}).

For the MS MARCO data set, we set the maximum input sequence length to 128 as we did not find significant differences between using 128 and 512. Similarly, for the NQ data set we set it to $128+256=384$ which shows performance similar to using 512.
The batch size is set to 32 for both data sets. 
Notice that the batch size here specifies the number of lists, and the effective query-document pairs in a batch is $36 \times 32 = 1152$.
We use a constant learning rate of $1 \times 10^{-4}$ during fine-tuning.
For the MS MARCO data set, we fine-tune our models for 50k steps. 
For the NQ data set, most of our models are fine-tuned for 100k steps, except the ones fine-tuned with pointwise cross-entropy loss, which achieves the best performance at 25k steps.
For Poly1 loss, we simply set $\epsilon=1$ for all of our experiments. 

The implementation is based on T5X, a T5 implementation\footnote{\url{https://github.com/google-research/t5x}} in JAX.
All the ranking losses are implemented in Rax, a learning-to-rank framework\footnote{\url{https://github.com/google/rax}}~\cite{rax} in JAX.

%% file: results.tex
\section{Results}
\label{sec:results}

\begin{table*}[!t]
    \centering
    \caption{
        Comparing ranking performances of different ranking models. 
        The best performances are bolded. 
        Results with $^{\dagger}$ are statistically significantly ($p \leq 0.05$) better than monoT5.
    }
    \scalebox{0.8}{%
      
        \begin{tabular}{cc|ccc|ccc}
        \toprule
         \multirow{2}{*}{Model} & \multirow{2}{*}{Loss} & \multicolumn{3}{c|}{MS MARCO} & \multicolumn{3}{c}{NQ} \\
         & & MRR@10 & NDCG@5 & NDCG@10 & MRR@10 & NDCG@5 & NDCG@10 \\
        \midrule
                  BERT & PointCE & 0.3867 & 0.4127 & 0.4530 & 0.5157 & 0.5515 & 0.5733 \\
                  BERT & Softmax & 0.3928 & 0.4173 & 0.4580 & 0.5213 & 0.5566 & 0.5791 \\
                 \midrule
                  monoT5 & Generation & 0.4156 & 0.4448 & 0.4843 & 0.5406 & 0.5861 & 0.6079 \\
                 \midrule
                  \multirow{4}{*}{RankT5-EncDec} 
                         & PointCE   & 0.4209 & 0.4496 & 0.4895$^{\dagger}$ & 0.5403 & 0.5833 & 0.6079 \\ 
                         & Pair      & 0.4177 & 0.4456 & 0.4852 & 0.5574$^{\dagger}$ & 0.5957$^{\dagger}$ & 0.6198$^{\dagger}$ \\ 
                         & Softmax   & 0.4278$^{\dagger}$ & 0.4573$^{\dagger}$ & 0.4960$^{\dagger}$ & 0.5687$^{\dagger}$ & \textbf{0.6068}$^{\dagger}$ & \textbf{0.6291}$^{\dagger}$ \\ 
                         & Poly1     &  \textbf{0.4343}$^{\dagger}$ & \textbf{0.4640}$^{\dagger}$ & \textbf{0.5025}$^{\dagger}$ & 0.5647$^{\dagger}$ & 0.6032$^{\dagger}$ & 0.6255$^{\dagger}$ \\  \midrule
                  \multirow{4}{*}{RankT5-Enc} 
                         & PointCE   & 0.4216$^{\dagger}$ & 0.4509$^{\dagger}$ & 0.4888 & 0.5441  & 0.5851 & 0.6099 \\ 
                         & Pair       & 0.4206 & 0.4513$^{\dagger}$ & 0.4891$^{\dagger}$ & 0.5595$^{\dagger}$  & 0.5980$^{\dagger}$ & 0.6215$^{\dagger}$ \\ 
                         & Softmax    & 0.4305$^{\dagger}$ & 0.4582$^{\dagger}$ & 0.4987$^{\dagger}$ & 0.5620$^{\dagger}$  & 0.6018$^{\dagger}$ & 0.6231$^{\dagger}$ \\
                         & Poly1      & 0.4296$^{\dagger}$ & 0.4586$^{\dagger}$ & 0.4970$^{\dagger}$ & \textbf{0.5689}$^{\dagger}$  & \textbf{0.6068}$^{\dagger}$ & 0.6279$^{\dagger}$ \\
        \bottomrule
        \end{tabular}
      
    }
    \label{tab:loss_comparison}
\end{table*}

\paragraph{Overall comparison.}
We compare the performances of our proposed rankers with different model structures and different losses. 
We also train BERT rankers~\cite{tfrbert} on the same data sets with the PointCE loss and the Softmax loss respectively. The BERT rankers are initialized from BERT-Large checkpoint, which has similar capacity as T5-Large encoders.
We also train the monoT5 model~\cite{nogueira2020T5ranking} on our data sets and report the performance.
Results are presented in Table~\ref{tab:loss_comparison}.

It can be observed that our proposed RankT5 variants achieve substantial improvement in ranking performance.
On the MS MARCO data set, our best model improves MRR@10, NDCG@5 and NDCG@10 by more than +1.8\% compared to monoT5. 
On the NQ data set, the best performance from our proposed models also outperforms monoT5 by +2.8\% in terms of MRR@10 and +2.0\% in terms of NDCG@5 and NDCG@10. 
This result verifies that by optimizing T5 models 
with ranking losses, we can significantly improve their text ranking performance.

We can also observe that some ranking losses consistently outperform others. 
For example, listwise losses like Softmax and Poly1 consistently outperform PointCE and Pair on both data sets.
For the encoder-decoder model structure on MS MARCO and the encoder-only model structure on NQ, the Poly1 loss performs significantly ($p \leq 0.05$) better than the Softmax loss on all metrics.

We also verify that the initial checkpoint of T5 shows advantage over the initial checkpoint of BERT. 
RankT5-Enc fine-tuned with the Softmax loss outperforms the BERT ranker fine-tuned with the Softmax loss on both data sets with a large margin (+3.7\% and +4.0\% respectively).
The same can be observed for both models fine-tuned with the PointCE loss, which is also reported in~\cite{nogueira2020T5ranking}. 
This result demonstrates the importance of choosing a good initial checkpoint and justifies the necessity to adapt T5 for ranking.

We do not find a consistent winner between the encoder-decoder (EncDec) and the encoder-only (Enc) model structure.
It seems that when using the same ranking losses, both model structures deliver similar performances. 
A possible explanation is that the T5 decoder is less important when the model is fine-tuned for text ranking tasks with sufficient training data.

\begin{figure}[t]
  \centering
  \subfigure[MS MARCO]{
    \label{subfig:size_msmarco}
    \includegraphics[width=0.46\columnwidth]{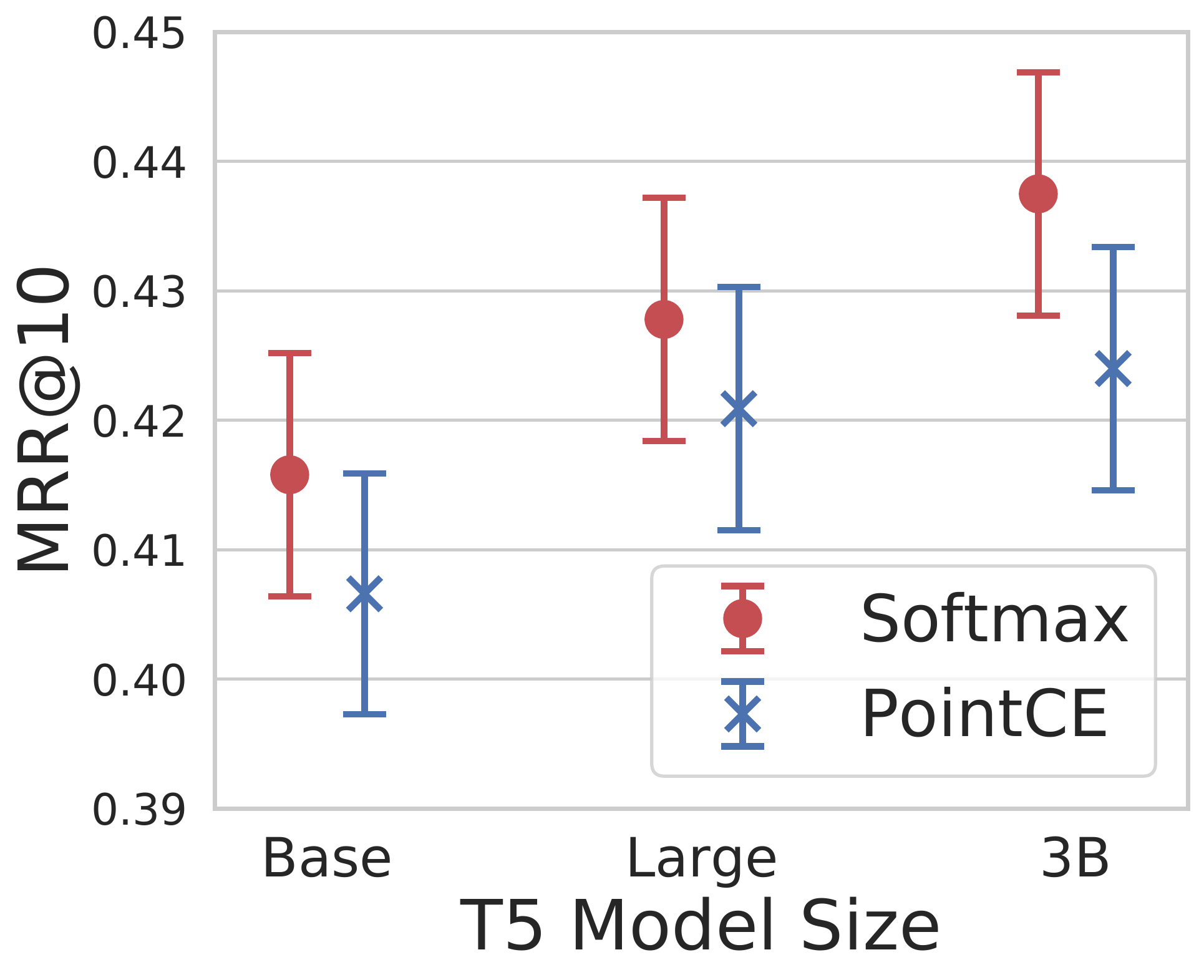}
  }
  \subfigure[NQ]{
    \label{subfig:size_nq}
    \includegraphics[width=0.46\columnwidth]{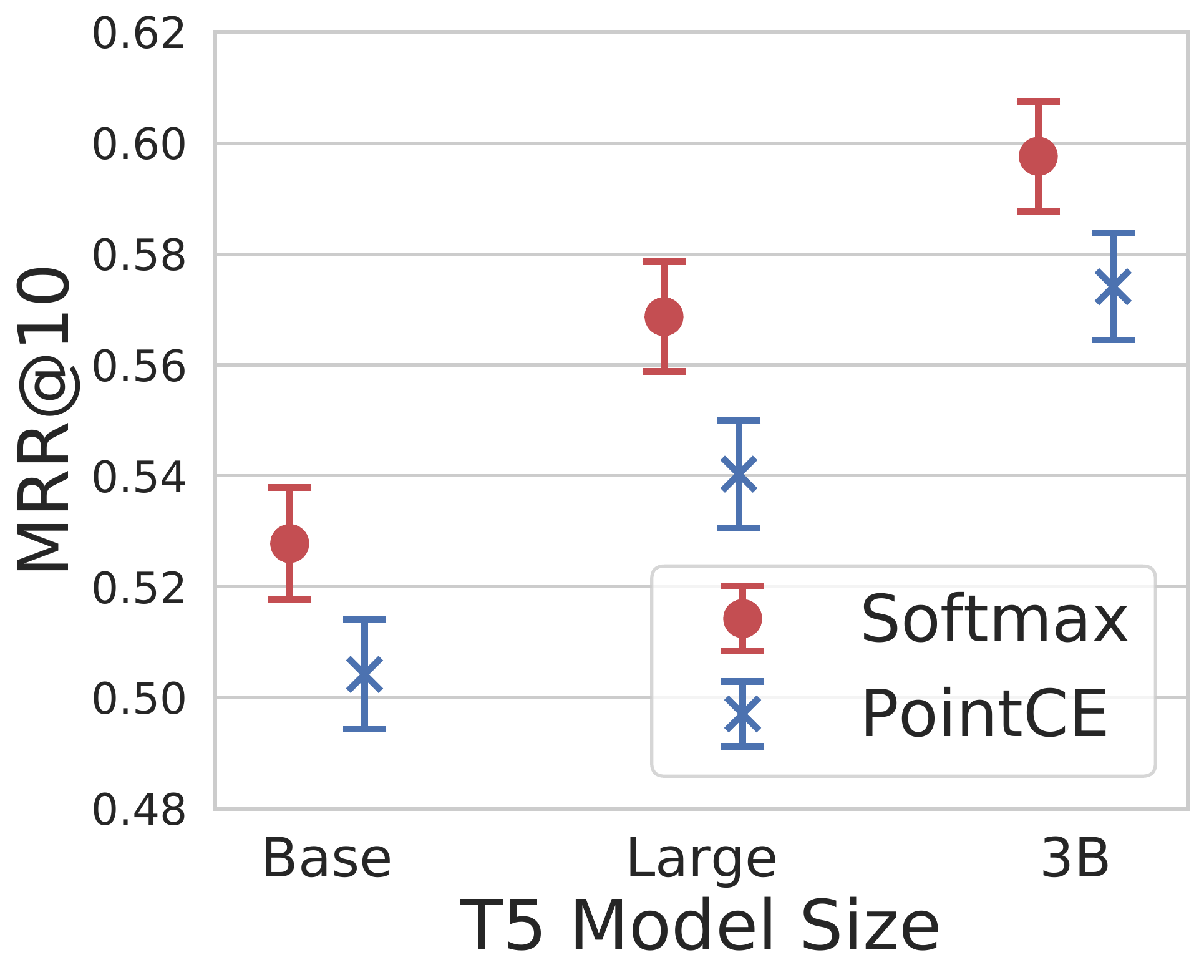}
  }
  \caption{\label{fig:size}
  Comparing performances with different T5 model sizes. Ranking models are RankT5-EncDec fine-tuned with the Softmax and the PointCE loss. The performance is measured by MRR@10 with 95\% confidence interval.
  }
  \label{fig:model_size}
\end{figure}

\paragraph{Model size comparison.}
We examine how the T5 model size affects the ranking performance.
We fine-tune the RankT5-EncDec model with the Softmax and the PointCE loss with different sizes of T5 model checkpoints, which are ``Base'', ``Large'' and ``3B''. 
We evaluate the model performances on both data sets measured by MRR@10.
Results are plotted in Figure~\ref{fig:model_size}.

The first observation is that the performance consistently improves when we use larger model size. 
On the NQ data set, the largest model (3B) outperforms the smallest model (Base) with a nearly +7\% gap. 
It would be worth trying even larger language models developed recently~\cite{chowdhery2022palm} in a similar way for ranking tasks. 

Another observation is that the models fine-tuned with the Softmax loss consistently outperform the ones fine-tuned with the PointCE loss (all statistically significant with $p \leq 0.05$) and the gaps remain relatively stable across different model sizes. This might suggest that extra signals captured by using the appropriate ranking loss cannot be compensated by simply using larger pretrained checkpoints.

\begin{figure}[t]
  \centering
  \subfigure[MS MARCO]{
    \label{subfig:size_msmarco}
    \includegraphics[width=0.46\columnwidth]{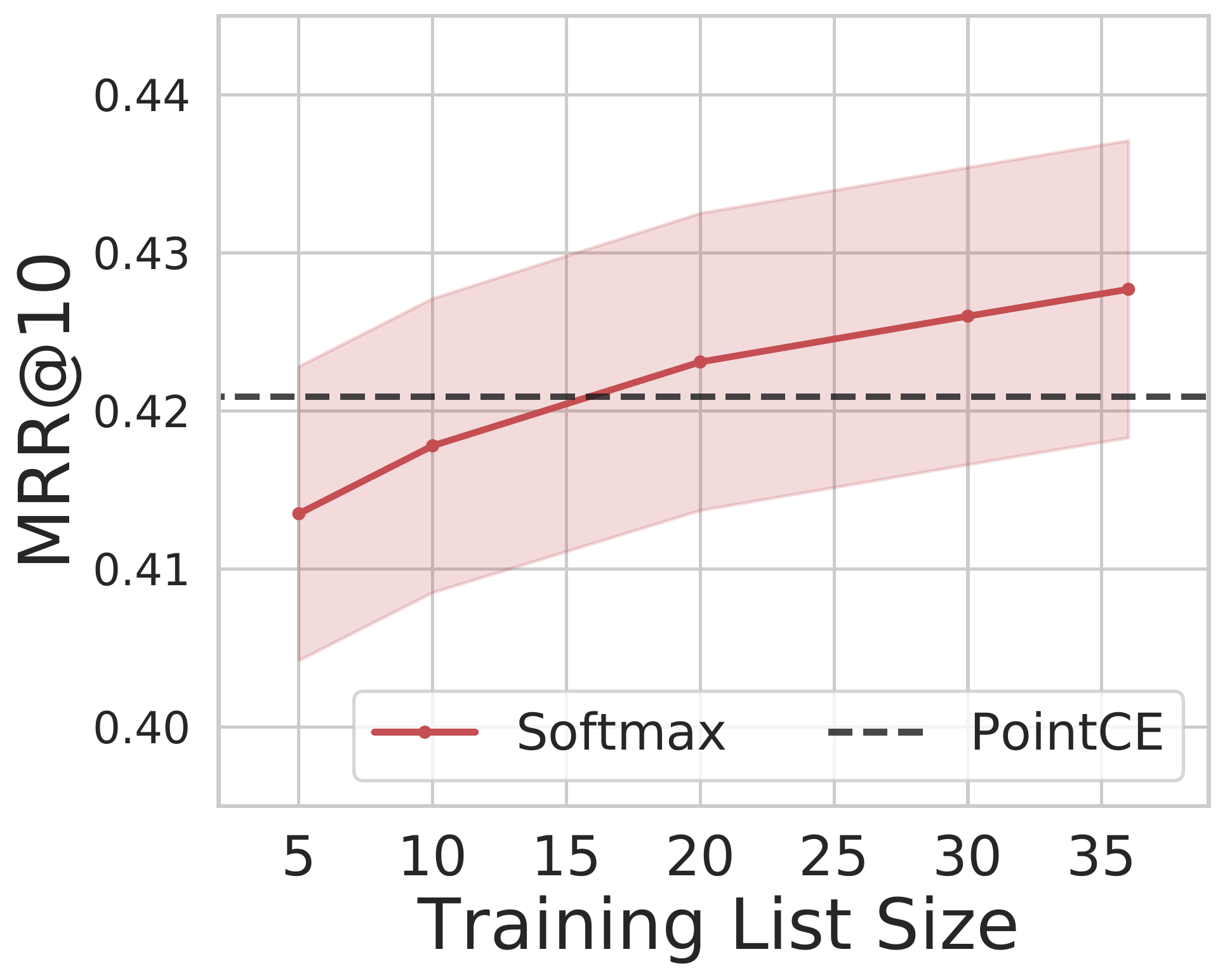}
  }
  \subfigure[NQ]{
    \label{subfig:size_nq}
    \includegraphics[width=0.46\columnwidth]{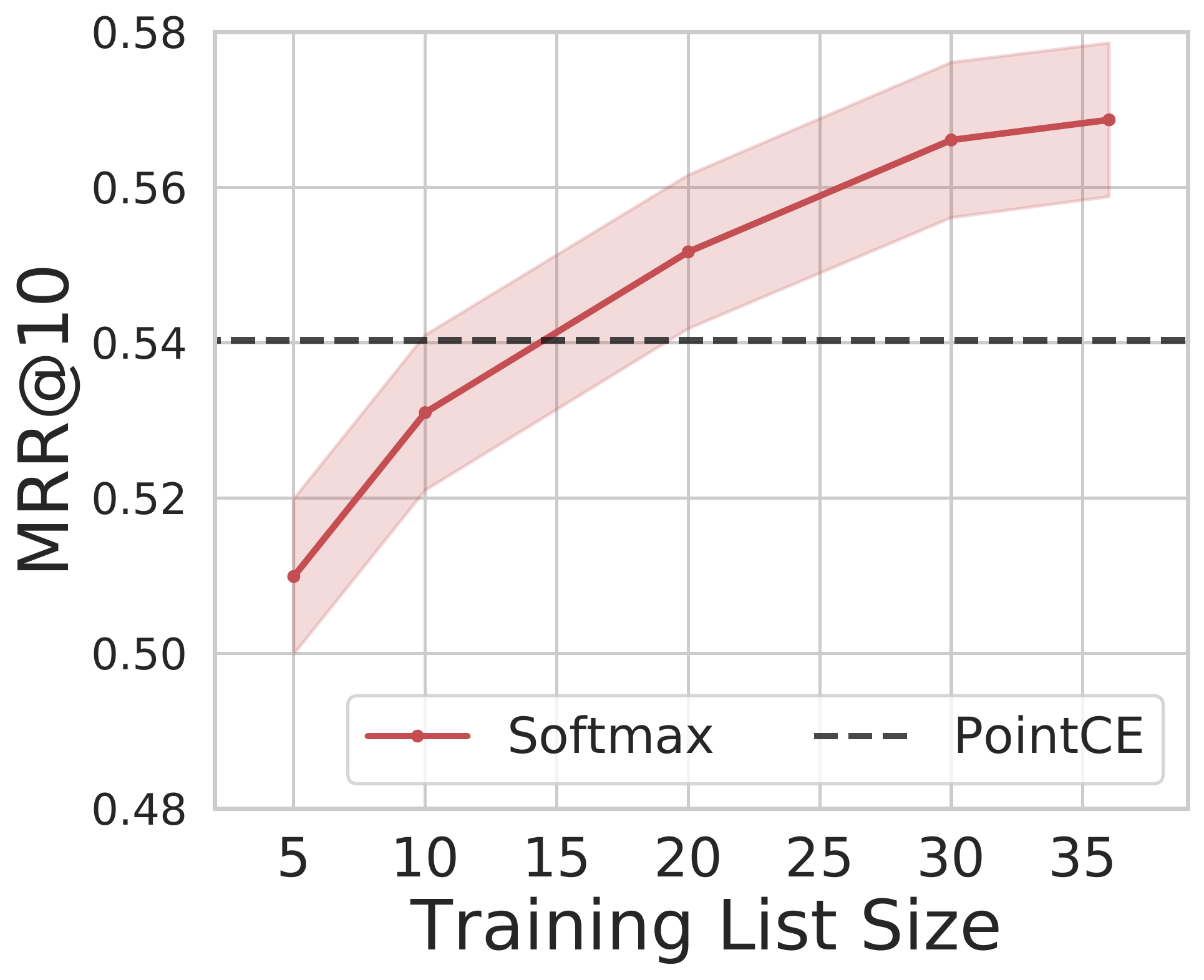}
  }
  \caption{\label{fig:list_size}
  Comparing performances with different training list sizes. Ranking models are RankT5-EncDec fine-tuned with Softmax loss. Similar models trained on PointCE loss are also plotted as the dashed line. The performance is measured by MRR@10.
  }
\end{figure}

\paragraph{Training list size.}
Another potential factor with a great impact on the ranking performance is the training data. 
We investigate how the size of each list in the training data affects the ranking performance, particularly when ranking losses like the Softmax loss is adopted. 
We prepare training data with different list sizes $m$, where each list still contains $1$ positive document and $(m-1)$ negative documents as described in Section~\ref{sec:exp}. 
In particular, we use $m$ from $\{5, 10, 20, 30, 36\}$ where 36 is the upper limit due to the hardware constraint.
Then we fine-tune the RankT5-EncDec model with the Softmax loss using each training list size and compare their performances. 
Results are illustrated in Figure~\ref{fig:list_size}.

It shows that on both data sets, the model performances improve with larger list sizes. 
For the NQ data set, the improvement is more significant but seems to saturate when the list size is over 30. 
For the MS MARCO data set, there also seems to be a consistent uptrend of the performance when the list size increases. 
These results suggest that using a large enough list size is crucial to obtain the optimal performance when fine-tuning with losses like the Softmax loss. 
We also plot the performances of ranking models fine-tuned with the PointCE loss for reference.
Notice that the list size needs to be at least around 20 to 30 for the Softmax loss to beat the PointCE loss.

\begin{table}[!t]
    \centering
    \caption{Zero-shot performance comparison. Ranking models are RankT5-Enc fine-tuned on the MS MARCO data set with different losses. Performances are measured by NDCG@10. The best performance for each data set is bolded.}
    {
        \small
        \begin{tabular}{c|cc}
        \toprule
        Data set      & PointCE & Softmax\\
        \midrule
        TREC-COVID    & 0.7522 & \textbf{0.8071} \\
        BioASQ        & 0.5346 & \textbf{0.5635} \\
        NFCorpus      & 0.3263 & \textbf{0.3810} \\
        NQ            & 0.5959 & \textbf{0.6142} \\
        HotpotQA      & \textbf{0.7126} & 0.7100 \\
        FiQA-2018     & 0.4156 & \textbf{0.4450} \\
        Signal-1M     & 0.3153 & \textbf{0.3200} \\
        ArguAna       & 0.2232 & \textbf{0.3300} \\
        Touché-2020   & \textbf{0.4594} & 0.4401 \\
        Quora         & 0.8221 & \textbf{0.8309} \\
        DBPedia       & 0.4345 & \textbf{0.4422} \\
        SCIDOCS       & \textbf{0.1821} & 0.1806 \\
        FEVER         & \textbf{0.8352} & 0.8316 \\
        Climate-FEVER & 0.2062 & \textbf{0.2152} \\
        SciFact	      & 0.7493 & \textbf{0.7499} \\
        \midrule
        Average       & 0.5024 & \textbf{0.5241} \\
        \bottomrule
        \end{tabular}
    }
    \label{tab:zero_shot}
\end{table}

\paragraph{Zero-shot results.}
We also compare the zero-shot performances of our ranking models fine-tuned with different ranking losses.
We use a subset of BEIR~\cite{thakur2021beir} data sets\footnote{Notice that the NQ data set in BEIR is has a different corpus and query set from the NQ data set we used earlier.} with easily accessible corpus.
We take our RankT5-Enc ranking models fine-tuned on the MS MARCO data set with the PointCE loss and the Softmax loss respectively, and apply these models to rerank top-1000 documents returned by BM25~\cite{robertson2009probabilistic}.
Table~\ref{tab:zero_shot} summarizes ranking performances measured by NDCG@10.

The ranking model fine-tuned with the Softmax loss outperforms the PointCE loss on 11 out of the 15 data sets.
On average, the Softmax loss achieves more than +2.1\% NDCG@10 (statistically significant with $p \leq 0.05$). 
These results indicate that using the Softmax loss produces ranking models that generalize better to out-of-domain data. 

In particular, using the Softmax loss achieves larger improvement on data sets with drastically different corpus. 
For example, TREC-COVID, BioASQ and NFCorpus are all in the biomedical domain with a very different vocabulary from the web corpus of MS MARCO.
The average improvement on these corpora reaches +4.6\%. 
This might imply that fine-tuning the model with appropriate ranking losses can enforce the model to put less emphasis on memorization, and thus to better learn the abstract concept of ``relevance''~\cite{lafferty2003probabilistic}, regardless of what the underlying corpus is.

%% file: conclusion.tex
\section{Conclusion}
\label{sec:conclusion}
In this paper, we study to utilize pretrained T5 models on text ranking.
Unlike monoT5, which simply converts a text ranking problem into a T5-compatible token generation task, our paper presents two model variants that output numerical numbers based on the T5 architecture: an encoder-decoder model and an encoder-only model. 
We thereby propose fine-tuning T5 with ranking losses to maximize ranking metrics. 
We demonstrate significant improvement when ranking losses are used to fine-tune the T5 model on the MS MARCO and the NQ data sets. 
We further illustrate that such improvement can be maintained in the zero-shot setting on out-of-domain data sets.

Our paper opens up several potential directions for further exploration. 
First, more T5-based ranking model architectures can be studied. For example, one can study different pooling strategies in the encoder-only model structure.
Second, it is unclear whether there is any difference between the encoder-decoder and the encoder-only models when the training data is insufficient.
Third, it might be interesting to extend our training method into the pretraining stage to produce better pretrained models.

%% file: appendix.tex
\appendix

\section{Choice of Target Tokens and Input Sequence}
\label{sec:target_token}

In Section~\ref{subsec:model_structure}, our proposed model structure with the encoder-decoder structure adopts ``\TARGET'' as the target token and uses its unnormalized logit as the ranking score. 
We believe that the choice of target token does not affect the model performance when we have sufficient training data. 
We conduct experiments where we use ``\texttt{<extra\_id\_11>}'' or ``\TRUE'' as the target token. 
We also try to use the original input sequence format in~\cite{nogueira2020T5ranking} where the input sequence contains the ``Relevant:'' postfix:
\begin{align}
s_{ij} = \text{Query:}~q_i~\text{Document:}~d_{ij}~\text{Relevant:} \nonumber
\end{align}
and the target token is ``\TRUE''.
We fine-tune our EncDec model on the MS MARCO data set with the Softmax loss using different target tokens or input sequence formats. 
Results are shown in Table~\ref{tab:target_token}.

\begin{table}[h]
    \centering
    \caption{Comparison of different target token choices and input sequence formats. All models are RankT5-EncDec model fine-tuned with the Softmax loss. The performances are reported on the MS MARCO data set.
    }
    \scalebox{0.75}{%
        \begin{tabular}{@{}c@{\hskip4pt}|@{\hskip4pt}c@{\hskip4pt}|@{\hskip4pt}c@{\hskip4pt}c@{}}
        \toprule
        Input sequence & Target token  & MRR@10 & NDCG@10 \\
        \midrule
        \multirow{3}{*}{w/o ``Relevant:''}
                 & \TARGET                  &  0.4277 & 0.4960 \\ 
                 & \texttt{<extra\_id\_11>} &  0.4272 & 0.4951 \\ 
                 & \TRUE                    &  0.4250 & 0.4937 \\ \midrule
        w/ ``Relevant:'' 
                 & \TRUE                    &  0.4279 & 0.4953 \\ 
        \bottomrule
        \end{tabular}
    }
    \label{tab:target_token}
\end{table}

It can be observed that the model performances are not sensitive to the choice of target tokens. 
Whether or not to add the ``Relevant:'' postfix to the input sequence also does not seem to show any effect. 
These results seem to align with the observation in~\cite{nogueira2020T5ranking} that the choice of target tokens does not change the performance when there is sufficient training data.

\section{Pooling Strategies Comparison for RankT5-Enc}
\label{sec:enc_pooling}

In Section~\ref{subsec:model_structure}, our proposed model structure with the encoder-only structure uses a pooling layer.
There could be different ways to instantiate the pooling layers. 
The first one is to only take the embedding vector of the first token:
\begin{align}
\label{eq:first_pooling}
 \text{Pool}_{\text{First}}([\be_1, \cdots, \be_l]) = \be_1 
\end{align}
Another common strategy is to take the mean of the embedding vectors of all the tokens:
\begin{align}
\label{eq:mean_pooling}
 \text{Pool}_{\text{Mean}}([\be_1, \cdots, \be_l]) = \frac{1}{l}\sum_{i=1}^{l}\be_i
\end{align}

We conduct experiments comparing these pooling strategies on MS MARCO, where the RankT5-Enc model is fine-tuned with the Softmax loss. 
Results are summarized in Table~\ref{tab:pooling}.

\begin{table}[h]
    \centering
    \caption{Comparison of different pooling strategies. All models are RankT5-Enc model fine-tuned with the Softmax loss. The performances are reported on the MS MARCO data set.
    }
    {
        \small
        \begin{tabular}{c|cc}
        \toprule
        Pooling  & MRR@10 & NDCG@10 \\
        \midrule
        First       &  0.4305 & 0.4987 \\ 
        Mean        &  0.4308 & 0.4993 \\ 
        \bottomrule
        \end{tabular}
    }
    \label{tab:pooling}
\end{table}

These results suggest that different pooling strategies do not generate much different results.
This is probably because even just the first token is already attended by all the other token with the transformer structure.
Moreover, in our experiments, there are sufficient training data to fine-tune the model.

\begin{table*}[h!]
    \centering
    \caption{
        Comparing ranking performances of different ranking models with additional metrics. 
        The best performances are bolded. 
        Results with $^{\dagger}$ are statistically significantly ($p \leq 0.05$) better than monoT5.
    }
    \scalebox{0.8}{%
      
        \begin{tabular}{cc|ccc|ccc}
        \toprule
         \multirow{2}{*}{Model} & \multirow{2}{*}{Loss} & \multicolumn{3}{c|}{MS MARCO} & \multicolumn{3}{c}{NQ} \\
         & & NDCG & MAP & Recall@5 & NDCG & MAP & Recall@5 \\
        \midrule
                  BERT & PointCE & 0.5180 & 0.3932 & 0.5528 & 0.6164 & 0.5228 & 0.6849 \\
                  BERT & Softmax & 0.5212 & 0.3978 & 0.5548 & 0.6215 & 0.5276 & 0.6895 \\
                 \midrule
                  monoT5 & Generation & 0.5421 & 0.4206 & 0.5945 & 0.6378 & 0.5434 & 0.7334 \\
                 \midrule
                  \multirow{4}{*}{RankT5-EncDec} 
                         & PointCE & 0.5462$^{\dagger}$ & 0.4260$^{\dagger}$ & 0.5973 & 0.6406 & 0.5457$^{\dagger}$ & 0.7417$^{\dagger}$ \\
                         & Pair & 0.5438 & 0.4229 & 0.5911 & 0.6531$^{\dagger}$ & 0.5629$^{\dagger}$ & 0.7392 \\
                         & Softmax & 0.5520$^{\dagger}$ & 0.4326$^{\dagger}$ & 0.6059$^{\dagger}$ & \textbf{0.6616}$^{\dagger}$ & 0.5740$^{\dagger}$ & 0.7480$^{\dagger}$ \\
                         & Poly1 & \textbf{0.5569}$^{\dagger}$ & \textbf{0.4383}$^{\dagger}$ & \textbf{0.6143}$^{\dagger}$ & 0.6586$^{\dagger}$ & 0.5701$^{\dagger}$ & 0.7458$^{\dagger}$ \\ \midrule
                  \multirow{4}{*}{RankT5-Enc} 
                         & PointCE & 0.5467$^{\dagger}$ & 0.4269$^{\dagger}$ & 0.5973 & 0.6431 & 0.5496$^{\dagger}$ & 0.7380 \\
                         & Pair & 0.5464$^{\dagger}$ & 0.4256 & 0.6029$^{\dagger}$ & 0.6547$^{\dagger}$ & 0.5650$^{\dagger}$ & 0.7418$^{\dagger}$ \\
                         & Softmax & 0.5539$^{\dagger}$ & 0.4347$^{\dagger}$ & 0.6052$^{\dagger}$ & 0.6564$^{\dagger}$ & 0.5674$^{\dagger}$ & \textbf{0.7467}$^{\dagger}$ \\
                         & Poly1 & 0.5536$^{\dagger}$ & 0.4344$^{\dagger}$ & 0.6059$^{\dagger}$ & 0.6615$^{\dagger}$ & \textbf{0.5744}$^{\dagger}$ & 0.7457$^{\dagger}$ \\
        \bottomrule
        \end{tabular}
      
    }
    \label{tab:loss_comparison_more_metrics}
\end{table*}

\section{Ranking Performances with Additional Metrics}

We provide some additional metrics to compare the performance of our RankT5 models with other baselines. 
The metrics we reported in Table~\ref{tab:loss_comparison} focus more precision, which emphasizes ensuring the top positions of ranked lists are relevant documents.
In this appendix we also report metrics like NDCG without cutoff or mean average precision (MAP), which would also penalize models if any relevant document is ranked too low in the list. Specifically, we report model performance measured by NDCG, MAP and Recall@5 in Table~\ref{tab:loss_comparison_more_metrics}. 

The conclusion is similar to Table~\ref{tab:loss_comparison}.
Our proposed RankT5 models still outperform other baselines significantly on these metrics.
Models fine-tuned with Softmax or Poly1 still achieve better performance than models fine-tuned with pointwise or pairwise losses.
These results are not surprising as both data sets have very few documents labeled for each query. 
Actually, for both data sets, more than 94\% of queries in the dev partition have only 1 document labeled as relevant.

\section{Additional Experiments on Model Size Comparison}
\label{sec:additional_model_size}

\begin{figure}[h]
  \centering
  \subfigure[MS MARCO]{
    \label{subfig:size_msmarco}
    \includegraphics[width=0.46\columnwidth]{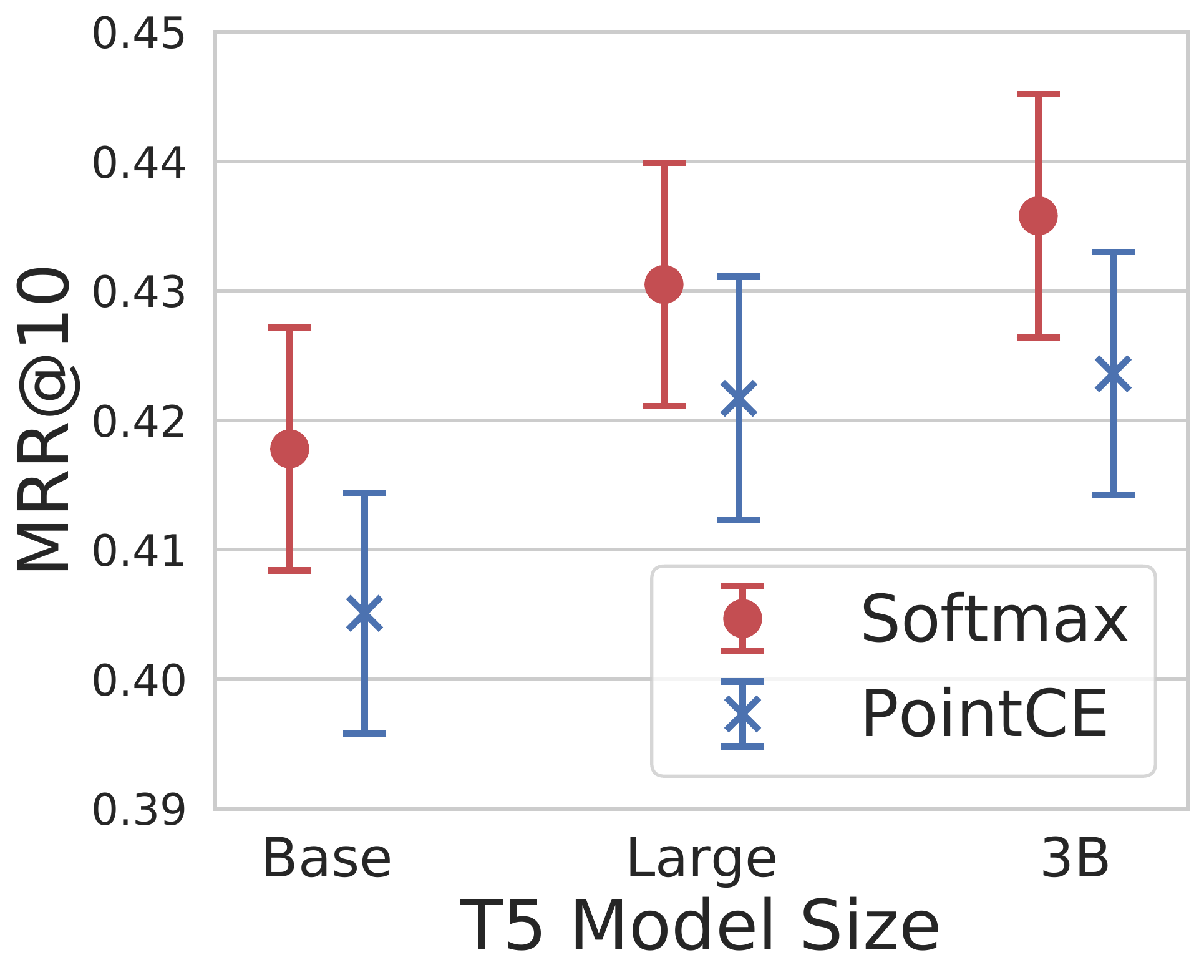}
  }
  \subfigure[NQ]{
    \label{subfig:size_nq}
    \includegraphics[width=0.46\columnwidth]{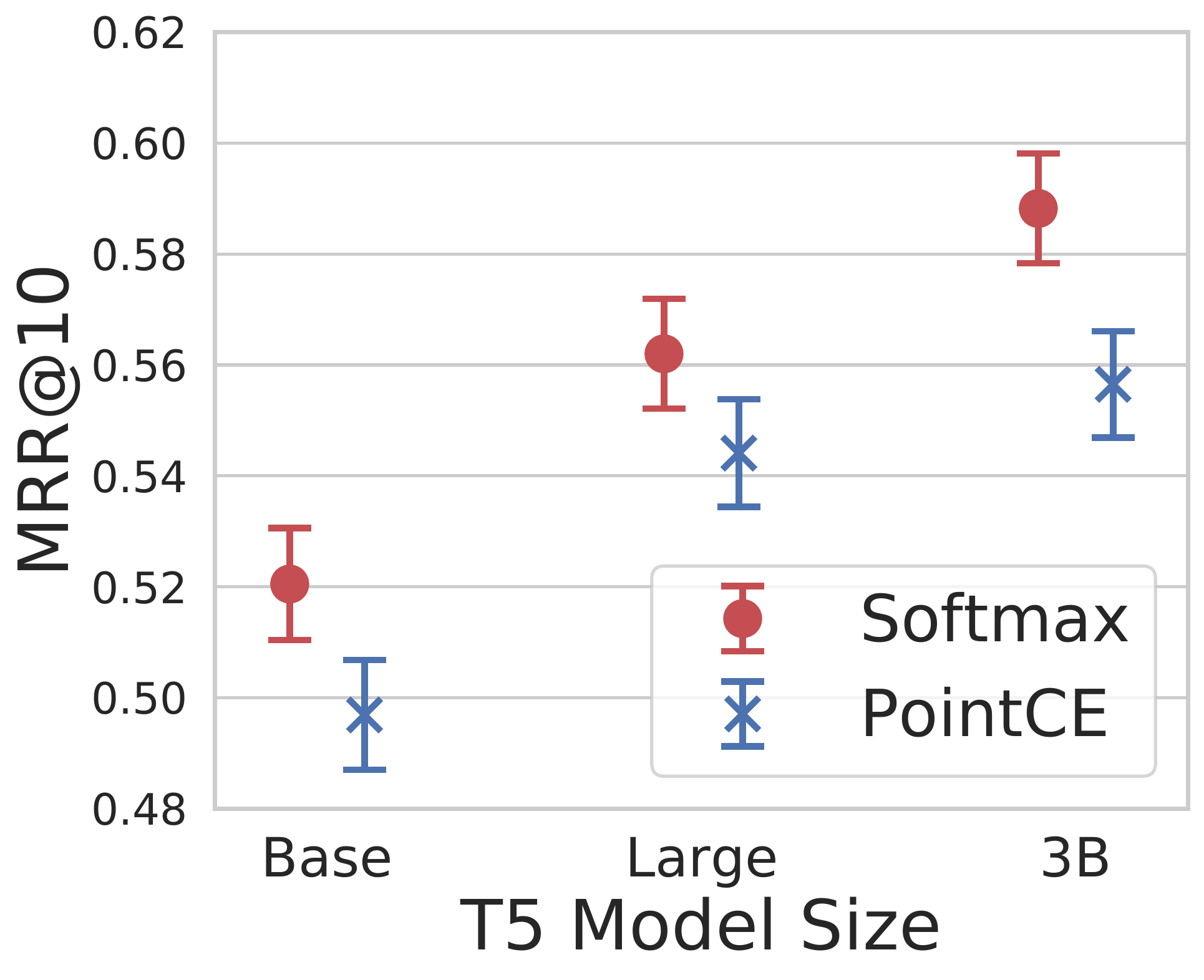}
  }
  \caption{\label{fig:size}
  Comparing performances with different T5 model sizes. Ranking models are RankT5-Enc fine-tuned with the Softmax and the PointCE loss. The performance is measured by MRR@10 with 95\% confidence interval.
  }
  \label{fig:model_size_enc}
\end{figure}

We report additional experiments of model size comparisons on RankT5-Enc models.
Similar to the experiments in Section~\ref{sec:results}, we fine-tune the RankT5-Enc model with the Softmax and the PointCE loss with T5 model sizes of ``Base'', ``Large'' and ``3B'' and plot their MRR@10. 
Results are plotted in Figure~\ref{fig:model_size_enc}.

The results are similar to RankT5-EncDec results. The performance increases with larger model sizes and the gap between the Softmax loss and the PointCE loss remains similar.

\section{Additional Zero-Shot Experiments}
\label{sec:additional_zero_shot_exp}

We also conduct the zero-shot experiments using RankT5-Enc fine-tuned with different model sizes.
All the models are fine-tuned using the Softmax loss. 
We compare the models with size ``Base'', ``Large'' and ``3B''. 
The results are shown in Table~\ref{tab:zero_shot_model_size}.

\begin{table}[h]
    \centering
    \caption{Zero-shot performance comparison. Ranking models are RankT5-Enc fine-tuned on the MS MARCO data set using the Softmax loss with different model sizes. Performances are measured by NDCG@10.}
    {
        \small
        \begin{tabular}{c|ccc}
        \toprule
        
        Data set & Base & Large & 3B \\
        \midrule
        TREC-COVID    & 0.7896 & 0.8071 & 0.8237 \\ 
        BioASQ        & 0.5627 & 0.5635 & 0.5795 \\ 
        NFCorpus      & 0.3731 & 0.3810 & 0.3990 \\ 
        NQ            & 0.5731 & 0.6142 & 0.6471 \\ 
        HotpotQA      & 0.7269 & 0.7100 & 0.7536 \\ 
        FiQA-2018     & 0.4132 & 0.4450 & 0.4932 \\ 
        Signal-1M     & 0.2909 & 0.3200 & 0.3191 \\ 
        ArguAna       & 0.3094 & 0.3300 & 0.4069 \\ 
        Touché-2020   & 0.4449 & 0.4401 & 0.4869 \\ 
        Quora         & 0.8138 & 0.8309 & 0.8192 \\ 
        DBPedia       & 0.4373 & 0.4422 & 0.4598 \\ 
        SCIDOCS       & 0.1760 & 0.1806 & 0.1918 \\ 
        FEVER         & 0.8295 & 0.8316 & 0.8487 \\ 
        Climate-FEVER & 0.2462 & 0.2152 & 0.2753 \\ 
        SciFact       & 0.7493 & 0.7499 & 0.7600 \\
        \midrule
        Average       & 0.5157 & 0.5241 & 0.5509 \\
        \bottomrule
        \end{tabular}
    }
    \label{tab:zero_shot_model_size}
\end{table}

It can be observed that on average the ranking model with larger size can also perform better on out-of-domain data sets in the zero-shot setting. 


We also compare the zero-shot performance of our RankT5-Enc model with publicly reported zero-shot performance of monoT5 from~\cite{rosa2022no}.
We only compare the performance with model size Base and 3B since these are the only sizes we have results from both models. 
The results can be found in Table~\ref{tab:zero_shot_monot5}.

\begin{table}[h]
    \centering
    \caption{Comparing zero-shot performance of RankT5-Enc and monoT5. RankT5-Enc is
    fine-tuned on the MS MARCO data set using the Softmax loss. Performances are measured by NDCG@10.}
    {
        \small
        \begin{tabular}{@{}c@{\hskip3pt}|cc|cc@{}}
        \toprule
        \multirow{2}{*}{Data set}  & \multicolumn{2}{c|}{Base} & \multicolumn{2}{c}{3B} \\
         & monoT5 & RankT5 & monoT5 & RankT5 \\
        \midrule
        TREC-COVID & 0.7775 & 0.7896 & 0.7948 & 0.8237 \\
        BioASQ & 0.5240 & 0.5627 & 0.5740 & 0.5795 \\
        NFCorpus & 0.3570 & 0.3731 & 0.3837 & 0.3990 \\
        NQ & 0.5674 & 0.5731 & 0.6334 & 0.6471 \\
        HotpotQA & 0.6950 & 0.7269 & 0.7589 & 0.7536 \\
        FiQA-2018 & 0.4136 & 0.4132 & 0.5137 & 0.4932 \\
        Signal-1M & 0.2771 & 0.2909 & 0.3140 & 0.3191 \\
        ArguAna & 0.1321 & 0.3094 & 0.2876 & 0.4069 \\
        Touché-2020 & 0.2773 & 0.4449 & 0.2995 & 0.4869 \\
        Quora & 0.8230 & 0.8138 & 0.8407 & 0.8192 \\
        DBPedia & 0.4195 & 0.4373 & 0.4777 & 0.4598  \\
        SCIDOCS & 0.1649 & 0.1760 & 0.1970 & 0.1918  \\
        FEVER & 0.8018 & 0.8295 & 0.8495 & 0.8487  \\
        Climate-FEVER & 0.2451 & 0.2462 & 0.2802 & 0.2753  \\
        SciFact & 0.7356 & 0.7493 & 0.7773 & 0.7600  \\
        \midrule
        Average & 0.4807 & 0.5157 & 0.5321 & 0.5509  \\
        \bottomrule
        \end{tabular}
    }
    \label{tab:zero_shot_monot5}
\end{table}

Results show that RankT5 fine-tuned with Softmax on average outperforms the monoT5 with both model sizes.
It again provides evidence that fine-tuning ranking models with listwise ranking losses can help the model generalize better to out-of-domain data, similar to our observation in Table~\ref{tab:zero_shot}.

\section{Limitations}
\label{sec:limitations}

One limitation of this work is that both data sets we used only have binary relevance labels. 
There are other text ranking data sets with graded relevance labels (e.g., 0 to 4), but they either are not publicly available or do not have a large enough training set.
Different fine-tuning strategies (training data construction, ranking losses, learning rate etc.) might need to be developed for our proposed models to achieve optimal performances on such data sets.

Another limitation is that we only focus on T5.
There are other sequence-to-sequence language models like GPT-2 with different model structures (where the encoder is not available). 
How to apply techniques proposed in this work to those models remains to be further discussed.